# Links between traumatic brain injury and ballistic pressure waves originating in the thoracic cavity and extremities

Amy Courtney, PhD
Department of Physics, United States Military Academy, West Point, NY 10996
Amy_Courtney@post.harvard.edu

Michael Courtney, PhD
Ballistics Testing Group, P.O. Box 24, West Point, NY 10996
Michael_Courtney@alum.mit.edu

**Abstract:**
Identifying patients at risk of traumatic brain injury (TBI) is important because research suggests prophylactic treatments to reduce risk of long-term sequelae. Blast pressure waves can cause TBI without penetrating wounds or blunt force trauma. Similarly, bullet impacts distant from the brain can produce pressure waves sufficient to cause mild to moderate TBI. The fluid percussion model of TBI shows that pressure impulses of 15-30 psi cause mild to moderate TBI in laboratory animals. In pigs and dogs, bullet impacts to the thigh produce pressure waves in the brain of 18-45 psi and measurable injury to neurons and neuroglia. Analyses of research in goats and epidemiological data from shooting events involving humans show high correlations ($r > 0.9$) between rapid incapacitation and pressure wave magnitude in the thoracic cavity. A case study has documented epilepsy resulting from a pressure wave without the bullet directly hitting the brain. Taken together, these results support the hypothesis that bullet impacts distant from the brain produce pressure waves that travel to the brain and can retain sufficient magnitude to induce brain injury. The link to long-term sequelae could be investigated via epidemiological studies of patients who were gunshot in the chest to determine whether they experience elevated rates of epilepsy and other neurological sequelae.

**Introduction**

Even without penetrating brain injury or blunt force trauma, patients with blast injury can experience traumatic brain injury (TBI) and are at heightened risk for long term sequelae from neurological damage caused by the blast pressure wave. [1-5] Evidence suggests that some patients with gunshot wounds may similarly sustain mild to moderate TBI. Identifying these patients is important because research suggests that some patients with mild to moderate TBI also suffer long-term neurological sequelae. [6-8]

Recent and ongoing research suggests low-cost and low-risk prophylactic treatments could be broadly applied. [9] For example, magnesium salts appear to reduce the size of the TBI lesion, and cyclosporine A (an inhibitor of mitochondrial transition pore opening and resulting apoptosis) has been shown to attenuate axonal injury, also decreasing the resultant lesion volume. At minimum, current recommendations for treatment of concussion, including a period of rest and attention to nutrition may be indicated. [10]

For patients at greater risk for long-term effects of mild to moderate TBI, other therapies may also be indicated. The Defense and Veterans Brain Injury Center (DVBIC), as well as other collaborators in the US federal TBI research programme initiated in 1996 are conducting clinical trials on a number of therapies (such as substance P inhibitors) as well as developing brief assessment tools for mild TBI.

In its publication Explosions and Blast Injuries: A Primer for Clinicians, the US Centers for Disease Control and Prevention (CDC) instructs:

<u>Primary blast waves can cause concussions or mild traumatic brain injury (MTBI) without a direct blow to the head … symptoms of concussion and post-traumatic stress disorder can be similar … Blast injuries are not confined to the battlefield. They should be considered for any victim exposed to an explosive force.</u> [11]

In this review of neurological and wound ballistics literature, it becomes clear that a ballistic pressure wave from distant bullet impact can induce a concussive-like effect in humans, causing acute neurological symptoms. Taken together, these studies also suggest that this effect may lead to long-term neurological sequelae in some patients. Epidemiological studies are suggested that may establish the value of routinely evaluating gunshot patients for mild or moderate TBI.

**Fluid percussion model**
The lateral fluid percussion (LFP) model of TBI has been used for more than 15 years to investigate anatomical and physiological mechanisms of traumatic brain injury. [12] In the LFP model, a brief pressure pulse is applied directly to the brain cortex of a laboratory animal by injecting a small volume of saline into the closed cranial cavity, producing a brief (20 ms) pressure transient. [13]

In a study applying a pressure wave directly to the brain of laboratory rats using the LFP technique, Toth et al. [13] report both instantaneous incapacitation and cellular damage:

<u>The delivery of the pressure pulse was associated with brief (120-200 sec) transient traumatic unconsciousness (as assessed by the duration of suppression of the righting reflex).</u>

Histology and immunocytochemical experiments demonstrated that the moderate LFP injury led to a loss of neurons from the hilus of the dentate gyrus of the hippocampus, and that this loss persisted months after the injury. However, many of the basket-like cells in the granule cell layer that were initially damaged survived the insult.

Injury to the hilar neurons occurred immediately due to the pressure wave on the large dentate neurons and not due to recruitment of active physiological processes (such as impact-induced increase in glutamate release). The physiological consequence of this neuronal loss is a decrease in the feed-forward activation of $GABA_A$ receptor-mediated inhibition. These results of LFP experiments agree with post-mortem examinations of head-injured patients that often reveal selective damage to the hippocampus and particularly to the dentate hilus. [14-16]

Using the LFP model of TBI, investigators quantified transient pressure levels associated with mild, moderate, and severe TBI, showing that mild and moderate injury levels occur with pressure levels in the 15-30 psi range. In fact, pressure waves near 30 psi caused immediate incapacitation in laboratory animals in the study by Toth et al. [13]

**CNS damage from pressure waves transmitted from bullet hits to extremities**
The pressure levels that produce mild to moderate TBI occurred in pigs and dogs that were shot in the thigh with high-energy projectiles. [17-19] Suneson et al. implanted high-frequency pressure transducers in the abdomen, neck, and brain of the test animals to measure pressures generated by distant missile impact at these locations. [18, Fig. 1] Transient pressure levels in the 18-45 psi range were transmitted to the brain. These pressures are comparable to pressures that produced incapacitation and long-term neural damage in LFP experiments.

In early tests, Suneson et al. [17] observed bursts of high frequency pressure waves in the abdomen and brain of anesthetized pigs. Brief apneic periods of a few seconds in length were also observed within the first minute after injury. Histological observations were limited to 'minor damage' to the blood-brain and blood-nerve barriers.

Subsequent experiments [18,19] revealed no gross damage or hemorrhage in large peripheral nerves or in the brain. Structural changes to the myelin sheaths, cytoskeleton and endoplasmic reticulum were microscopically apparent. In addition, chromatolysis was observed in many Purkinje cells in the cerebellum, and structural changes were observed in neurons of the hippocampus. The authors concluded that these effects were caused by high frequency pressure waves that were transmitted to the brain from the distant (0.5m) point of origin.

In order to localize CNS damage from high-energy missile extremity impact, and to look for a possible biomarker for brain damage in traumatic stress disorder, Wang et al. [20] performed similar experiments in dogs. Cerebrospinal fluid (CSF) samples were drawn from the cisterna magna via a small, previously implanted catheter 2, 4, and 6 hours after injury. Brain tissue was prepared and examined at 8 hours after injury. Myelin basic protein (MBP, a nervous system specific protein that reflects the extent of demyelination and amount of lost CNS tissue in pathologic processes) was measured in the CSF and fresh tissue samples.

Gross and light microscopic examination did not reveal significant damage to the tissues. Electron microscopic examination revealed changes localized to neurons in the hippocampus and hypothalamus, including nuclear membrane damage, decreased organelles, mitochondrial swelling, increased synaptic vescicles in the asymmetric synapse, and damage to the myelin sheath. MBP levels increased from the 6 to 8 hour samples; these levels positively correlated with MBP from the CSF and paralleled the ultrastructural damage that was observed.

The authors concluded:

<u>These findings corresponded well to the results of Suneson et al. [19], and confirmed that the distant effect exists in the central nervous system after a high-energy missile impact to an extremity.</u>

Moreover, localization of the damage to the hippocampus and (when the pressure wave is strong enough) the hypothalamus agrees with the results of LFP experiments. The results of these independent studies agree that a pressure wave generated by a high-energy missile impact at a distant location can reach the brain and can be strong enough to cause neuronal damage.

**Concussion and incapacitation from the ballistic pressure wave**
Not all bullet impacts produce a pressure wave strong enough to cause neurological symptoms. The likelihood of a bullet impact remote from the brain leading to rapid incapacitation has been demonstrated to increase with the local pressure wave magnitude. [21, 22] An appropriate mathematical model relating rapid incapacitation to pressure wave magnitude can quantify this effect. The value of this analysis lies in identifying the magnitude of pressure wave applied to the thoracic cavity that may result in pressures to the brain sufficient to cause mild to moderate TBI.

The largest published data set quantifying handgun bullet incapacitation in humans reveals contributions from both the ballistic pressure wave and the size of the wound channel. A mathematical model was developed for this data set and employed the hypothesis that these mechanisms are independent. [21] Functional forms for the probability of incapacitation by each mechanism were combined by the rules of probability to derive an empirical model for the total incapacitation probability.

The model fit the data well, with a correlation coefficient of r = 0.939. Independent contributions of each mechanism reveal that some pistol loads have pressure wave contributions that dominate the likelihood of rapid incapacitation. Most centerfire rifle and shotgun wounds to the chest involve pressure wave levels sufficient to contribute to rapid incapacitation.

The largest published data set quantifying handgun bullet incapacitation in animal test subjects (goats) shows that average incapacitation time correlates most strongly with ballistic pressure wave magnitude. [22] A plot of the average incapacitation time for different handgun loads versus the local (thoracic) peak pressure wave magnitude is shown in figure 1.

A model with the proper limiting behavior for the average incapacitation time incapacitation (AIT) as a function of the peak pressure wave magnitude, p, is:

$$AIT(p) = 10s\sqrt{\frac{p_0}{p}}, \quad (1)$$

where $p_0$ is the characteristic pressure wave that gives an average incapacitation time of 10 seconds. Performing a least-squares fit gives $p_0$ = 482 psi with a standard error of 1.64 s and a correlation coefficient of r = 0.91. This result suggests that the pressure wave effect begins to be noticeable at about 500 psi to the chest and that rapid incapacitation is likely to occur in about half of those experiencing about 1000 psi to the chest. The pressure wave produced by a specific load is a function of the local rate of energy transfer.

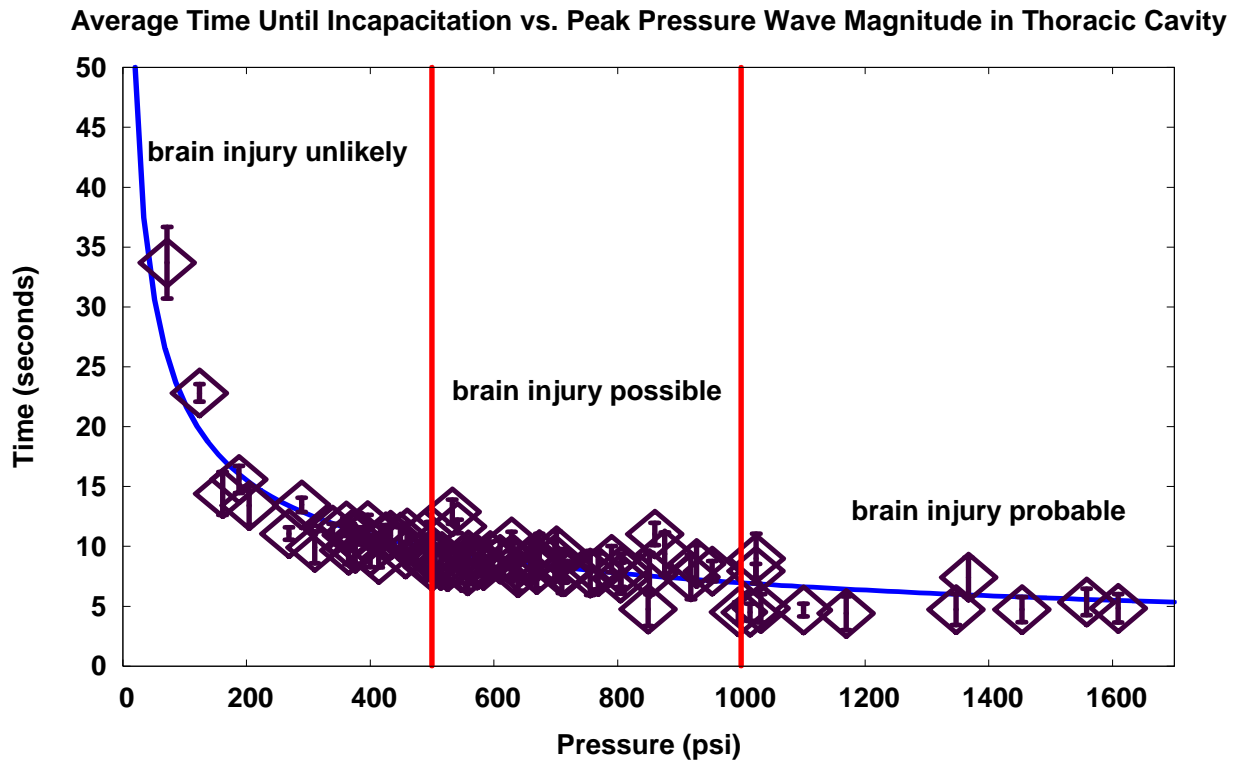

*Figure 1: A plot of average incapacitation time vs. pressure wave magnitude in the thoracic cavity, along with best-fit model.* [22]

Analysis of cumulative incapacitation probability as a function of time reveals both fast (t < 5 s) and slow (t > 5 s) incapacitation mechanisms. Since loss of consciousness due to loss of blood requires at least five seconds, [23] incapacitation in fewer than five seconds appears to have a neurological basis and result from the pressure wave. Indeed, the fast incapacitation mechanism can be accurately modeled as a function of local peak pressure wave magnitude. The slow incapacitation mechanism, then, is presumably due to blood loss via damaged vascular tissue.

The clinical implications of these analyses are that about half of patients who have sustained a gunshot wound to the chest that produced a local pressure wave of 1000 psi or greater are likely to have experienced the rapid (neurological) incapacitation effect and may have experienced mild to moderate TBI.

**Case study**
Treib et al. [24] reported a case study of a World War II veteran who, in 1941, received a bullet wound near but not directly impacting the brain. The patient experienced acute epileptic symptoms and the symptoms reappeared nearly 50 years later with secondary generalization.

A 7.62 mm bullet from a Russian Tokarev military pistol entered the patient's head just inferior to the nose, penetrated below the base of the cranium and lodged left of the second cervical vertebra. The brain itself was not directly struck. The patient suffered from focal seizures in the field hospital, and seizures ceased after a few years under medication. However, in 1990 seizures returned accompanied by secondary generalization.

This pistol is known to have muzzle velocities on the order of 1500 fps, and a muzzle energy of approximately 430 ft-lbs. With a penetration depth estimated at 8", this yields a local pressure wave magnitude of 1027 psi. [22] Trieb et al. write:

We suspect that the so-called hydrodynamic effect [pressure wave] of this high-velocity bullet caused an indirect trauma to the brain. This case shows that symptomatic epilepsy can occur after a penetrating head injury, without direct injury to brain tissue by a missile. [24]

**Summary and indications for further research**

The literature in this review provides evidence that some bullet impacts to locations distant from the brain can cause pressure waves sufficient to cause mild to moderate TBI. The fluid percussion model of TBI shows that pressure impulses of 15-30 psi cause mild to moderate TBI in laboratory animals. Experiments in pigs and dogs show that bullet impacts to the thigh can produce pressure waves in the brain of 18-45 psi and measurable injury to neurons and neuroglia. These experiments suggest that the hippocampus is the most susceptible to remote injury by ballistic pressure wave, and that the hypothalamus can also be damaged when the pressure wave is strong enough. Analyses of research in goats and epidemiological data from shooting events involving humans show high correlations (r > 0.9) between rapid (neurological) incapacitation and pressure wave magnitude (in the thoracic cavity).

It is widely believed that concussion is related to mild TBI [10,15,25-28] and that the loss of consciousness due to concussion can be correlated with heightened risk of long-term neurological sequelae. For example, Vanderploeg et al. [8] found that MTBI patients who experienced altered consciousness in motor vehicle accidents exhibited deficiencies in subtle aspects of complex attention and working memory as well as impaired tandem gait an average of eight years postinjury.

In a one-year follow-up study of patients with MTBI, Stalnacke et al. [7] found that about half of the patients experienced frequent persistent symptoms, disabilities and low levels of life satisfaction, though only about 5% remained on sick leave from employment. While few case studies have been published in which epilepsy and other serious neurological sequelae manifest decades after TBI, it is recognized by the neuropsychological community and the CDC that the symptoms of long-term sequelae of TBI can be similar to those of post-traumatic stress syndrome (PTSS) and that efforts should be made to determine whether TBI may have contributed to a patient's long-term neurological difficulties. [2,6,11]

Since management guidelines exist for mild to moderate TBI and since some low-cost and low-risk treatments are available [9-11,26,27], it would be beneficial to identify patients who have suffered mild to moderate TBI secondary to gunshot wounding. The results presented above suggest that it may be beneficial to evaluate patients who may have experienced a pressure wave in the thoracic cavity of 1000 psi or greater. This would include loads transferring 500 ft-lb of energy over 10 in of penetration or 600 ft-lb of energy over 12 in of penetration, for example. This would not include loads from a .22 LR rifle or most handgun loads. Most bullets from bottleneck centerfire rifle cartridges that expand, tumble, or fragment in the first six inches of tissue do produce local pressure wave magnitudes above 1000 psi. The local pressure wave for a specific case can be estimated if the bullet and cartridge are known [21, 22] and we suggest this estimate be used as an indicator for evaluation, not a basis for diagnosis at this time.

Studies of patients with MTBI due to other causes suggest that a small percentage (perhaps 5%) experience serious long-term sequelae, while as many as half may experience milder yet measurable declines in some cognitive and motor functions. It would be beneficial to determine the incidence and severity of long-term sequelae of MTBI secondary to remote gunshot wounding. A study of data already available in trauma centre or military healthcare databases may be possible, or these databases may be useful for identifying patients for long-term follow-up.

Selection criteria should be guided by the results of the studies reviewed above to identify an appropriate cohort. To maximize the likelihood of identifying patients who have suffered MTBI secondary to gunshot wounding, we suggest including patients who have sustained a gunshot wound to the thoracic cavity and no direct injury to the CNS, and cases in which the local pressure wave produced by the missile is likely to have been 1000 psi or greater. At this level, it is likely that approximately half of the patients experienced rapid incapacitation due to concussive-type effects. (While analyses suggest that a pressure wave to the chest above 500 psi may affect the brain, the large number of data points required to detect a small effect make it an unreasonably low criterion for initial studies.)

## About the Authors

*Amy Courtney* currently serves on the faculty of the United States Military Academy at West Point. She earned a MS in Biomedical Engineering from Harvard University and a PhD in Medical Engineering and Medical Physics from a joint Harvard/MIT program. She has taught Anatomy and Physiology as well as Physics. She has served as a research scientist at the Cleveland Clinic and Western Carolina University, as well as on the Biomedical Engineering faculty of The Ohio State University.

*Michael Courtney* earned a PhD in experimental Physics from the Massachusetts Institute of Technology. He has served as the Director of the Forensic Science Program at Western Carolina University and also been a Physics Professor, teaching Physics, Statistics, and Forensic Science. Michael and his wife, Amy, founded the Ballistics Testing Group in 2001 to study incapacitation ballistics and the reconstruction of shooting events. www.ballisticstestinggroup.org